\documentclass[final,1p,times]{elsarticle} 
\usepackage{graphicx} 
\usepackage{amssymb} 
\usepackage{amsthm} 
\usepackage{lineno} 

\newcommand{\nc}[1]{\newcommand{#1}}
\nc{\ben}{\begin{enumerate}}
\nc{\een}{\end{enumerate}}

\nc{\bo}[1]{\mbox{\boldmath \( #1 \! \! \)  \unboldmath}}

\nc{\be}{\begin{eqnarray}}
\nc{\ee}{\end{eqnarray}}
\nc{\la}{\langle}
\nc{\ra}{\rangle}
\def\simge{\mathrel{%
       \rlap{\raise 0.511ex \hbox{$>$}}{\lower 0.511ex \hbox{$\sim$}}}}
\def\simle{\mathrel{
       \rlap{\raise 0.511ex \hbox{$<$}}{\lower 0.511ex \hbox{$\sim$}}}}
\nc{\Fsi}{F^{\bf 1}}
\nc{\Fad}{F^{\bf 8}}
\nc{\Fsy}{F^{\bf 6}}
\nc{\Fan}{F^{{\bf 3}^*}}
\nc{\Fm}{F^M}
\nc{\tr}{{\rm tr \,}}
\nc{\Tr}{{\rm Tr \,}}
\nc{\bx}{{\bf x}}
\nc{\by}{{\bf y}}
\nc{\bz}{{\bf 0}}

\journal{Nuclear Physics A} 
\begin{document} 

\begin{frontmatter} 


\title{Free energies of heavy quarks in full-QCD lattice simulations\\
 with Wilson-type quark action}

\author{Y.~Maezawa$^1$,
        S.~Aoki$^{2,3}$,
        S.~Ejiri$^4$, 
        T.~Hatsuda$^5$, 
        N.~Ishii$^5$,
        K.~Kanaya$^2$,
        H.~Ohno$^2$,
    and T.~Umeda$^6$\\
(WHOT-QCD Collaboration)}

\address{
$^1$
En'yo Radiation Laboratory, Nishina Accelerator Research Center, RIKEN, Wako 351-0198, Japan 
\\
$^2$
Graduate School of Pure and Applied Sciences, University of Tsukuba, Tsukuba, Ibaraki 305-8571, Japan
\\
$^3$
RIKEN BNL Research Center, Brookhaven National Laboratory, Upton, New York 11973, USA 
\\
$^4$
Physics Department, Brookhaven National Laboratory, Upton, New York 11973, USA 
\\
$^5$
Department of Physics, The University of Tokyo, Tokyo 113-0033, Japan 
\\
$^6$
Graduate School of Education, Hiroshima University, Hiroshima 739-8524, Japan
}

\begin{abstract} 
The free energy between a static quark and an antiquark is 
 studied by using the color-singlet Polyakov-line correlation 
 at finite temperature in lattice QCD with 2+1 flavors of 
 improved Wilson quarks.
 From the simulations on $32^3 \times 12$, 10, 8, 6, 4 lattices  in the high temperature phase, 
 based on the fixed scale approach,
 we find that, the heavy-quark free energies at short distance converge
  to the  heavy-quark potential evaluated from the Wilson loop at zero temperature, 
  in accordance with the expected insensitivity of short distance physics to the temperature.
 At long distance, the heavy-quark free energies approach to 
 twice the single-quark free energies, implying that the interaction between heavy quarks
 is screened.
 The Debye screening mass obtained from the long range behavior of the free energy
 is compared with the results of thermal perturbation theory.
\end{abstract} 

\end{frontmatter} 



\section{Introduction and the method}\label{intro}

Heavy-ion collision experiments running at RHIC
 have provided various information about the new state of matter, the quark-gluon plasma (QGP).
To remove theoretical uncertainties in understanding the nature of QGP,
 first principle calculations by lattice QCD are indispensable.
 So far, most lattice studies at finite temperature have been
 performed using staggered-type quark actions with the fourth-root trick, 
 whose theoretical basis is not fully established yet. 
Thus, a crosscheck with other actions such as the Wilson-type quark actions are
important to control and estimate systematic errors due to lattice discretization.

 We investigate thermodynamic properties of QGP using the improved Wilson quark action.
 Adopting the fixed scale approach developed in \cite{WHOT-U}, we perform 
 finite-temperature simulations \cite{Kanaya}, while the corresponding zero-temperature configurations 
 are taken from the results of the CP-PACS and JLQCD Collaborations \cite{CP_JL}.
 
At $T=0$, interaction between a static quark and an antiquark 
 can be studied by the heavy-quark potential evaluated from the Wilson loop operator.
 The resulting potential takes the Coulomb form at short distances 
 due to perturbative gluon exchange, while it takes 
 the linear form  at long distances due to confinement:
\be
 V(r) = - \frac{\alpha_0}{r} + \sigma r + V_0 
.
\label{eq:V}
\ee
For $T>0$, inter-quark interaction may be studied by the heavy-quark free energy
 $\Fsi (r,T)$
 evaluated from a Polyakov-line  correlation in the color-singlet channel 
  with Coulomb gauge fixing \cite{Nadkarni}:
\be
\Fsi (r,T) = - T \ln \la \Tr \Omega^\dagger (\bx) \Omega (\by) \ra
,
\hspace{7mm}
r=|\bx - \by| ,
\ee
where $\Omega(\bx) = \prod_{\tau=1}^{N_t} U_4(\bx,\tau)$
with $U_4(\bx,\tau)$ being the link variable in the temporal direction. 
At zero temperature, we expect $\Fsi(r,T=0) = V(r)$, while
at high temperature, we can extract the Debye screening mass $m_D$ from
\be
\Fsi(r,T) = - \frac{\alpha(T)}{r} e^{-m_D(T)\, r} + 2F_Q
.
\label{eq:mD}
\ee

 In the conventional fixed $N_t$ approach where $T$ is varied by changing the lattice spacing $a$,  
 $V(r)$ and $\Fsi(r,T)$ receive different 
renormalization at each $T$, and thus are usually adjusted by hand such that
 $V(r)$ and $\Fsi(r,T)$ coincide with each other at a short distance,
  assuming that the short distance properties are insensitive to the temperature. 
On the other hand, in the fixed scale approach, 
the temperature $T=(N_t a)^{-1}$ is varied by changing the temporal 
lattice size $N_t$ at fixed $a$ \cite{WHOT-U}.
Because the spatial volume and the renormalization factors are common to all temperatures, 
we can directly compare the free-energies at different temperatures
 without any adjustment. 
We show below that $\Fsi(r,T)$'s for different $T$ approach to $V(r)$ at short distances, 
 which proves the expected insensitivity of the short distance physics to the temperature.

\section{Results of the lattice simulations}

We employ the renormalization group improved gluon action and $2+1$ flavors of nonperturbatively 
 $O(a)$-improved Wilson quark actions.
 Zero-temperature configurations are given by the CP-PACS and JLQCD Collaborations \cite{CP_JL}
 at $m_{\pi}/m_{\rho}=0.6337(38)$ and $m_K/m_{K^\ast} = 0.7377(28)$.  
 Finite temperature simulations with the same parameters are performed 
  on $32^3 \times N_t$ lattices with $N_t=12$, 10, 8, 6 and 4, which 
  correspond to $T \sim 200$--700 MeV \cite{Kanaya}.
 The absolute scale is estimated from the Sommer parameter, $r_0=0.5$ fm.

\begin{figure}[bt]
  \begin{center}
    \includegraphics[width=77mm]{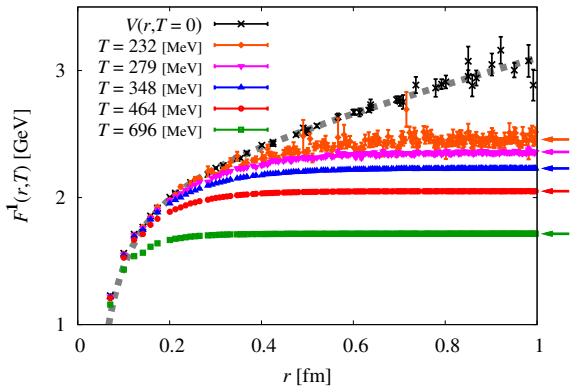}
    \vspace{-3mm}
    \caption{Heavy-quark free energies at various temperatures. 
    The heavy-quark potential at $T=0$ was calculated by the CP-PACS and JLQCD Collaborations \cite{CP_JL}.}
    \vspace{-3mm}
    \label{fig:1}
  \end{center}
\end{figure}

Figure \ref{fig:1} shows the results of the heavy-quark potential $V(r)$ at $T=0$ and
 the heavy-quark free energies $\Fsi(r,T)$ at various temperatures as functions of $r$.
At $T=0$, $V(r)$ shows Coulomb-like and linear-like behaviors at short and long distances, respectively.
A fit with Eq.~(\ref{eq:V}), as shown by the dashed line in Fig.~\ref{fig:1}, gives $\alpha_0 = 0.441$ and $\sqrt{\sigma} = 0.434$ GeV.

For $T>0$, we note that the heavy-quark free energies $\Fsi(r,T)$ at all 
temperatures converge to $V(r)$ at short distances.
This means that the short distance physics is insensitive to temperature.
As stressed above, unlike the case of the conventional fixed-$N_t$ approach
in which this insensitivity is assumed and  used to adjust the constant terms of $\Fsi(r,T)$,
 our fixed scale approach enabled us to directly confirm this theoretical expectation.

At large $r$, $\Fsi(r,T)$ departs from $V(r)$
 and eventually becomes flat due to  Debye screening.
  In  Fig.~\ref{fig:1}, the asymptotic values of $\Fsi(r,T)$ at long distance 
  are also compared with $2F_Q$ denoted by the arrows: Here  the thermal average
  of a single Polyakov-line is defined as $F_Q=-T \ln \la \Tr \Omega \ra$.
  We find that $\Fsi(r,T)$ converges to $2F_Q$ quite accurately at long distances.

\begin{figure}[bt]
  \begin{center}
    \includegraphics[width=62mm]{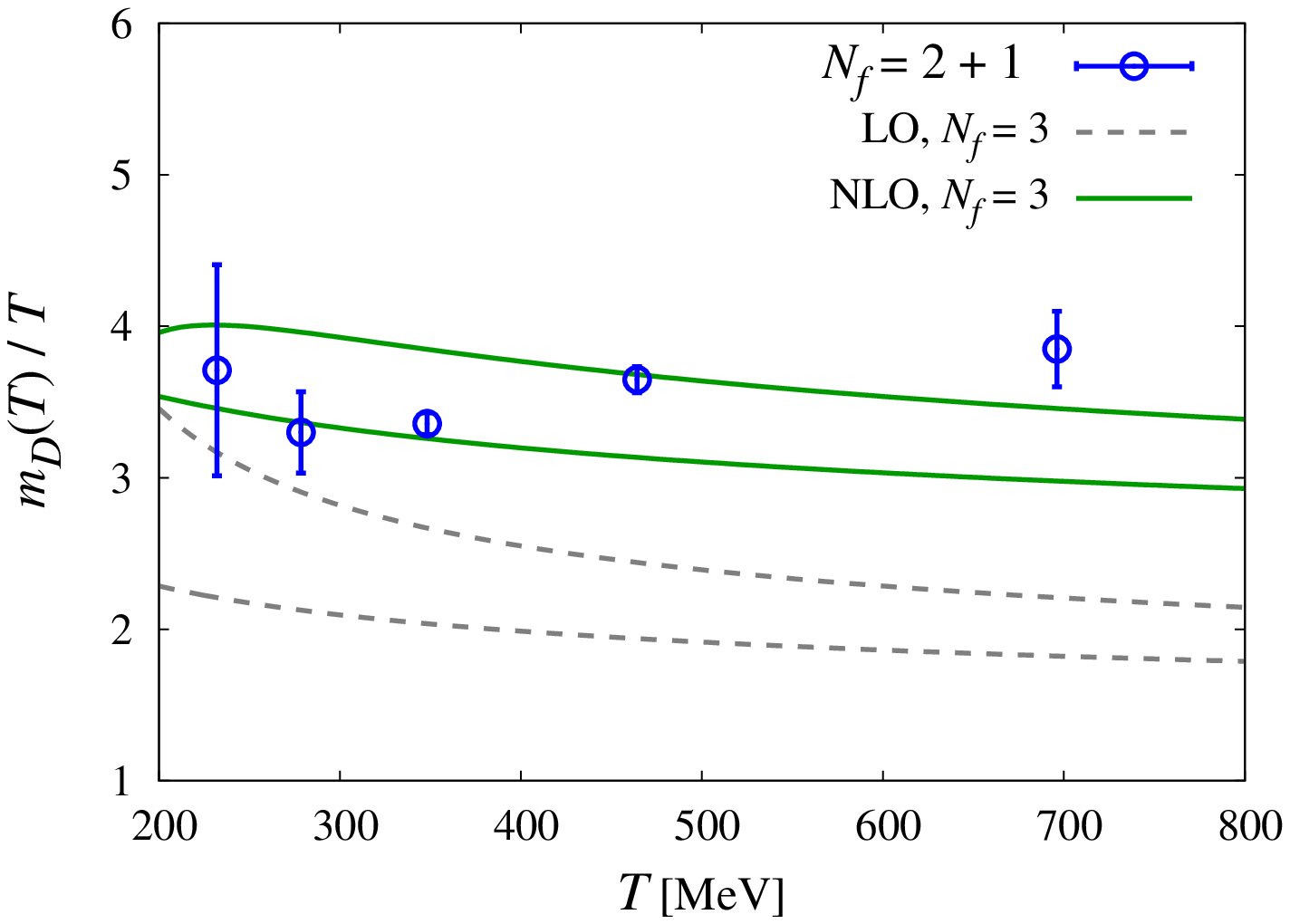} 
    \hspace{1mm}
    \includegraphics[width=62mm]{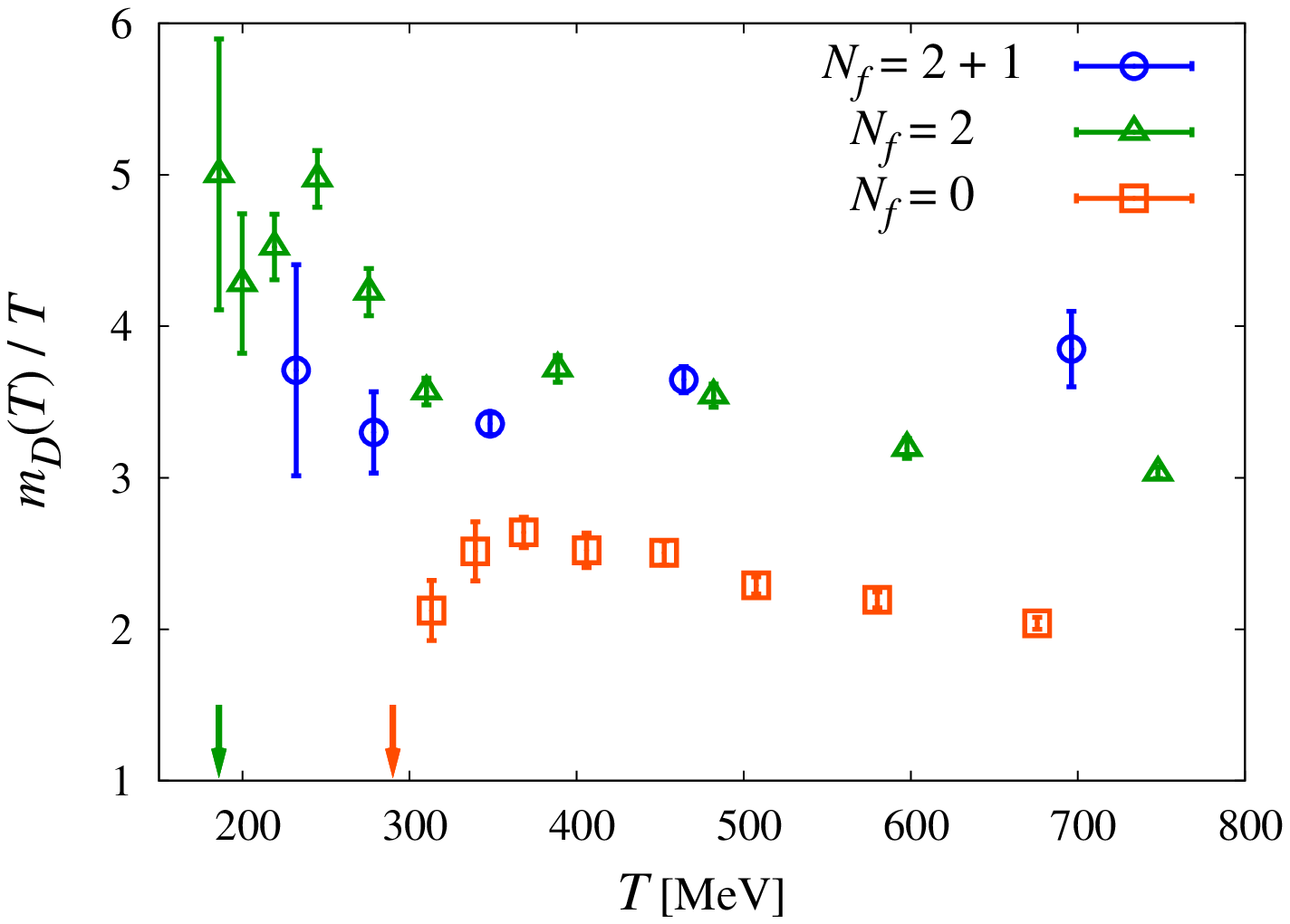}
    \vspace{-3mm}
    \caption{Left: Results of the Debye mass for $N_f=2+1$ obtained from the heavy-quark
     free energy with
   those of the leading order (LO) and next-to-leading order (NLO) thermal perturbation theory 
   for $N_f=3$. Right: Comparison among the Debye masses for $N_f=2+1$, $N_f=2$ \cite{WHOT-M}
   and $N_f=0$ \cite{WHOT-U}
   extracted from the heavy-quark free energy on the lattice.
   Arrows in the right panel indicate critical temperatures in $N_f=2$ and 0 cases.}
    \vspace{-3mm}
    \label{fig:2}
  \end{center}
\end{figure}

In order to extract the screening mass $m_D$,
 we fit $\Fsi(r,T)$ by the screened Coulomb form, Eq.~(\ref{eq:mD}).
The fit range is chosen to be 0.38 fm $\simle r \simle$ 0.57 fm.
The left panel of Fig.~\ref{fig:2} shows $m_D(T)/T$ which
 does not have strong dependence on $T$.
To make a  quantitative comparison to the result of the thermal perturbation theory,
 we define the  2-loop running coupling by
$g^{-2}_{\rm 2l} (\mu) 
=  \beta_0 \ln ( {\mu}/{\Lambda_{\overline{\rm MS}}} )^2 + 
\frac{\beta_1}{\beta_0} 
\ln \left[ \ln ( {\mu}/{\Lambda_{\overline{\rm MS}}} )^2 \right]$
with the QCD scale parameter 
 $\Lambda_{\overline{\rm MS}}^{N_f=3} = 260$ MeV \cite{Gockeler:2005rv},
 where we assume a degenerated $N_f=3$ case.
Then, the Debye mass  in the leading-order (LO) thermal perturbation theory is given by 
 $ {m_D^{\rm LO}(T)}/{T} = \sqrt{ 1 + N_f / 6 } \  g_{\rm 2l} (T)$ 
neglecting the quark mass effects.
A formula in the next-to-leading-order (NLO) is also 
available from the resummed hard thermal loop calculation \cite{Rebhan:1993az}:
\be
\frac{m_D^{\rm NLO}}{T} = 
\sqrt{ 1 + \frac{N_f}{6} } \
g_{\rm 2l}(T) \left[ 1 + 
g_{\rm 2l}(T) \frac{3}{2 \pi} \sqrt{\frac{1}{1+ N_f/6}}
\left(
\ln \frac{2 m_D^{\rm LO}}{m_{\rm mag}} - \frac{1}{2}
\right)
+ o(g^2)
\right]
,
\label{eq:m_D_NLO}
\ee
where $m_{\rm mag}(T) = C_m g^2(T) T$  is the magnetic screening mass.
Since the factor $C_m$ cannot be determined in perturbation theory due to the infrared problem,
  we adopt $C_m \simeq 0.482$ calculated in 
a quenched lattice simulation \cite{Nakamura:2003pu} as a typical value.
In Fig.~\ref{fig:2}(left), the dashed lines represent the LO results for $N_f=3$
and the bold lines represent the NLO results for $N_f=3$, 
for a range of the renormalization point $\mu = \pi T$--$3 \pi T$.
We find that 
the LO Debye mass does not reproduce the lattice data in magnitude,
while the NLO Debye mass is approximately 50 \% larger than the LO Debye mass
 and  shows a better agreement with the lattice data.

Finally we study the flavor dependence of $m_D$.
Fig.~\ref{fig:2}(right) shows $m_D$ in $N_f=2+1$ QCD (this work),
 that in $N_f=2$ QCD with an improved Wilson-quark action 
 at $m_{\pi}/m_{\rho} = 0.65$ \cite{WHOT-M}, and
 that in the quenched QCD ($N_f=0$) \cite{WHOT-U}.
Arrows on the horizontal axis indicate critical temperatures for
 $N_f=2$ ($T_c \sim 186$ MeV at $m_{\pi}/m_{\rho} = 0.65$) and 
  $N_f=0$ ($T_c \sim 290$ MeV).
We find that $m_D$ for $N_f = 2+1$ is comparable
 to that for $N_f=2$, whereas it is larger than that for $N_f=0$.
A similar result was obtained with a staggered-type quark action \cite{Petrov:2007ug}.

\section{Summary}

We studied the free energy between a static quark and an antiquark 
at finite temperature in lattice QCD with $2+1$ flavors of improved Wilson quarks 
on $32^3 \times 12$--4 lattices in the high temperature phase. 
We adopted 
the fixed scale approach which enables us to compare the free-energies at different
 temperatures directly without any adjustment of the overall constant. 
At short distances, the heavy-quark free energies, evaluated from the
 Polyakov-line correlations in the color-singlet channel,
 show universal Coulomb-like behavior common to that of the heavy-quark potential at zero temperature.
This is in accordance with the expected insensitivity of short distance physics to the temperature.
At long distances, the heavy-quark free energies approach
 to twice the single-quark free energies calculated from the thermal average
of a Polyakov-line. Also,
we  extracted the Debye screening mass $m_D(T)$ from the heavy-quark free energy
and found that the next-to-leading order  thermal perturbation theory is required 
to explain the magnitude of $m_D(T)$ on the lattice.
Comparison to the previous results at $N_f=2$ and $N_f=0$,
 shows that the dynamical light quarks have sizable effects on the value of $m_D(T)$.


We thank the members of CP-PACS and JLQCD Collaborations
 for providing us with the data at zero temperature.
This work is partially supported 
by Grants-in-Aid of the Japanese Ministry
of Education, Culture, Sports, Science and Technology, 
(Nos. 17340066, 18540253, 19549001, 20105001, 20105003, 20340047, 21340049). 
SE is supported by U.S. Department of Energy
(DE-AC02-98CH10886). 
This work is supported 
also by the Large-Scale Numerical Simulation
Projects of CCS/ACCC, Univ. of Tsukuba, 
and by the Large Scale Simulation Program of High Energy
Accelerator Research Organization (KEK) 
Nos.08-10 and 09-18.


\begin{thebibliography}{00} 

\bibitem{WHOT-U} T.~Umeda et al. (WHOT-QCD Collaboration),
 {\it Phys. Rev.} {\bf D79} (2009) 051501.

\bibitem{Kanaya}  K. Kanaya et al. (WHOT-QCD Collaboration), in these proceedings.

\bibitem{CP_JL} T.~Ishikawa et al., (CP-PACS and JLQCD Collaborations)
 {\it Phys. Rev.} {\bf D78} (2008) 011502.

\bibitem{Nadkarni} S. Nadkarni, {\it Phys. Rev.} {\bf D34} (1986) 3904.

\bibitem{Gockeler:2005rv}
  M.~Gockeler et al., {\it Phys. Rev.} {\bf D73} (2006) 01451.

\bibitem{Rebhan:1993az}
  A.~K.~Rebhan, {\it Phys. Rev.} {\bf D48} (1993) 3967.

\bibitem{Nakamura:2003pu}
  A.~Nakamura, T.~Saito and S.~Sakai, {\it Phys. Rev.} {\bf D69} (2004) 014506.

\bibitem{WHOT-M} Y.~Maezawa et al. (WHOT-QCD Collaboration),
 {\it Phys. Rev.} {\bf D75} (2007) 074501.

\bibitem{Petrov:2007ug}
  K.~Petrov et al.,  (RBC-Bielefeld Collaboration)
  {\it PoS} {\bf LAT2007}, (2007) 217.



\end{thebibliography}
\end{document}